\documentclass{article}
\usepackage{fullpage}
\usepackage{natbib}
\usepackage{graphicx}
\usepackage{amsmath,amssymb}
\usepackage{hyperref}
\usepackage{booktabs}

\usepackage{caption}
\usepackage{subcaption}
\usepackage{adjustbox} 
\usepackage{multirow} 
\usepackage{float}
\usepackage{ragged2e}

\usepackage{color}

\newcommand{\norm}[1]{\lVert #1 \rVert}
\newcommand{\R}{\mathbb{R}}

\usepackage{soul}

\usepackage{arydshln}

\begin{document}
\title{Improving Minimax Group Fairness in Sequential Recommendation}

\author{
    Krishna Acharya\thanks{Lead author. Work done while the author was an intern at Amazon.} \\
    Georgia Institute of Technology \\
    \texttt{krishna.acharya@gatech.edu}
    \and
    David Wardrope \\
    Amazon \\
    \texttt{dwardrop@amazon.co.uk}
    \and
    Timos Korres \\
    Amazon \\
    \texttt{korres@amazon.co.uk}
    \and
    Aleksandr Petrov\thanks{Work done while the author was an intern at Amazon.} \\
    University of Glasgow \\
    \texttt{a.petrov.1@research.gla.ac.uk}
    \and
    Anders Uhrenholt \\
    Amazon \\
    \texttt{akuhren@amazon.co.uk}
}

\date{}  
\maketitle          
\begin{abstract}
Training sequential recommenders such as SASRec with uniform sample weights achieves good overall performance but can fall short on specific user groups. One such example is popularity bias, where mainstream users receive better recommendations than niche content viewers. To improve recommendation quality across diverse user groups, we explore three Distributionally Robust Optimization(DRO) methods: Group DRO, Streaming DRO, and Conditional Value at Risk (CVaR) DRO. While Group and Streaming DRO rely on group annotations and struggle with users belonging to multiple groups, CVaR does not require such annotations and can naturally handle overlapping groups. In experiments on two real-world datasets, we show that the DRO methods outperform standard training, with CVaR delivering the best results. Additionally, we find that Group and Streaming DRO are sensitive to the choice of group used for loss computation. Our contributions include (i) a novel application of CVaR to recommenders, (ii) showing that the DRO methods improve group metrics as well as overall performance, and (iii) demonstrating CVaR's effectiveness in the practical scenario of intersecting user groups. Our code is available at \url{https://github.com/krishnacharya/sequentialrec-fairness}
\end{abstract}

\section{Introduction}
Recommender systems play a crucial role in user-facing services, from ecommerce \cite{Amazon-2decades,Alibaba-trans} to video\cite{YT-covington}, music streaming\cite{Spotify-seqcontext,deezer-trans} and social networks \cite{eksombatchai2018pixie,backstrom-socialnetwork}.
Among recommender systems, sequential recommenders specialize in next-item prediction, i.e., deciding what to recommend next based on the sequence of items a user has previously interacted with.
 Transformer-based models \cite{SasrecOG,gSAS,Bert4RecOG} excel here and have been deployed at scale\cite{Spotify-seqcontext,deezer-trans}. These model are usually trained using empirical risk minimization (ERM), i.e., by minimizing the loss uniformly for all examples in the training set.

While standard training (ERM) performs well overall, it can be suboptimal for specific user groups \cite{UserFairnessECIR,GroupFairnessWWW22,GroupFairnessSIGIR22}, as models often learn correlations that apply broadly but not within certain groups \cite{FairnessGooglePaper,FairnessRecSurvey}. A common example is popularity bias, where mainstream users receive better recommendations than minorities. In this work, we focus on minimax group fairness \cite{Minimax-fairness-Roth,SagawagDRO}, which minimizes the maximum loss across user groups, with the goal of improving metrics for all groups. This differs from traditional group fairness, which aims to equalize metrics across groups \cite{FairBespoke,FairDemMcAuley}. We consider two types of user groups: popularity-based groups and sequence length groups. These choices are motivated by prior findings that recommendation quality differs based on the popularity of items a user interacts with \cite{kowald2021-majority,kowald2020-majority,Popular-items-Abdollahpuri}, and across users with different sequence lengths \cite{volkovs2017dropoutnet,Netflix-coldstart}. Importantly, these user groups are chosen because they can be extracted for all datasets, and we expect the findings to generalize to other such attributes.

Previous work \cite{GoogleDRO10.1145/3485447.3512255} improves minimax group fairness for a non-sequential retrieval model using Group DRO (GDRO) \cite{SagawagDRO} and Streaming DRO (SDRO) \cite{GoogleDRO10.1145/3485447.3512255}. However, both methods require predefined group annotations and are also unable to handle users belonging to multiple groups. Additionally, attributes such as age, ethnicity are protected under GDPR\footnote{General Data Protection Regulation \url{https://gdpr-info.eu/}} and not disclosed by users \cite{opt-out}. As a result, GDRO and SDRO are inherently limited in achieving fairness for such groups; in fact, focusing only on known attributes can amplify bias \cite{chen2019fairness-unaware}.

In contrast, Conditional Value at Risk DRO (CVaR) \cite{cvarpaper} trains models to be robust to a wide range of shifts by identifying subsets of each training mini-batch with the highest losses and updating the model to minimize those. This approach does not require group annotations and dynamically up-weights the highest-loss examples in each mini-batch. In an orthogonal study, Singh et al. \cite{singh2020building} investigate safe reinforcement learning and apply a CVaR objective to optimize worst-case reward trajectories. In contrast, we focus on improving minimax fairness across user groups in the standard setting of training a sequential recommender (SASRec \cite{SasrecOG}).
In this paper, we evaluate the effectiveness of CVaR  DRO alongside Group DRO and Streaming DRO. Our contributions are: 
\begin{enumerate}
\item We demonstrate the effectiveness of Conditional Value at Risk DRO for SASRec, a transformer-based sequential recommender. To the best of our knowledge, this is the first application of CVaR DRO to recommender systems for addressing minimax group fairness.
\item Through an in-depth evaluation across group sizes, we show that the DRO methods improve NDCG scores on user groups while also achieving higher overall NDCG compared to standard training. Notably, CVaR, which requires no group information outperforms both group and streaming DRO.
\item We evaluate these methods in a practical scenario with users belonging to intersecting groups, a previously unexplored setup. Here too, the DRO methods outperform standard training, with CVaR often surpassing group and streaming DRO. We also highlight the sensitivity of the GDRO and SDRO results to the group selected for loss computation.
\end{enumerate}
Our experimental evidence suggests that practitioners should prioritize group-agnostic methods like CVaR over group-dependent approaches such as GDRO or SDRO. CVaR not only scales easily to multiple groups but also performs better. \\

The paper is organised as follows: Section \ref{sec:seqrec+methods} provides background on sequential recommenders and the training methods; 
Section \ref{sec:exp-groupsize-effect} contains experiments with varying group sizes on the Retailrocket dataset, Section \ref{sec:exp-intersecting-groups} then evaluates the methods with intersecting user groups; Section \ref{sec:conclusion} concludes the paper; Appendix \ref{app:ml1m-exp} repeats the experiments in Section \ref{sec:experiments-intro} on the Movielens-1M dataset.

\section{Sequential recommenders and training methods}
\label{sec:seqrec+methods}
In this section, we provide the necessary background on sequential recommenders and the training methods: standard training(empirical risk minimization); the class balanced (CB), inverse propensity weighted (IPW) baselines; and the three distributionally robust methods: Conditional Value at Risk(CVaR) DRO, group DRO and streaming DRO.
\subsection{Sequential recommenders}
In sequential recommendation we want to predict the next item for each user $u$ given $H_u = (h_u^1, \ldots h_u^n)$, a chronological sequence of $n$ items with which a user has interacted. Each $h_u^j$ is an item from the item catalogue $I$. There have been several approaches based on Recurrent Neural Networks \cite{GRU4Rec,LRURec} and Transformers \cite{SasrecOG,Bert4RecOG} towards this, but the main idea is similar: 1) Transform the user sequence $H_u$ to an embedding $\in \R^d$, 2) Multiply this embedding by a matrix of learnt item embeddings $W_I^\theta \in \R^{|I| \times d}$ to obtain \textit{scores} for each item, 3) The items with the highest scores are sent as candidates to  a downstream ranker.\\
In the context of SASRec, a transformer decoder model, a user sequence $H_u$ is first padded/truncated to a length $L$ of most recent items, combined with positional embeddings and then passed through the transformer. For detailed information about the SASRec architecture we refer the reader to \cite{SasrecOG}. The $j^{th}$ item in the sequence $h_u^j$ serves as a target for the previous $j-1$ items. Denoting the transformer by $f^\theta(h) \to \R^d$ a function from item history to embeddings, the logits and cross entropy loss for user $u$ at position $j$ are 
\[
Logits_{u, j}^\theta =W_I^\theta \times f^\theta([h_u^1, \ldots, h_u^{j-1}]) \in \R^{|I|} ~~~~~ \ell_{u,j}^\theta = \text{CE}(\text{SoftMax}(Logits_{u, j}^\theta), h_u^j)
\]
$\ell_u^\theta = \frac{1}{L}\sum_{j=1}^L \ell_{u,j}^\theta$ is defined as the average cross entropy loss for user $u$. 
Note that while the original SASRec \cite{SasrecOG} is setup as binary classification with the binary cross entropy loss, recent papers \cite{gSAS,dross,acceralated-softmax} demonstrate that applying softmax over all items followed by the cross entropy loss achieves state-of-the-art accuracy. We adopt this approach in our work.
\subsection{Standard training (ERM)} 
Standard training, formally known as Empirical Risk minimization(ERM), minimizes the average loss across users. 
The loss given a mini-batch of $B$ users:
\begin{align}
Loss_{ERM}^\theta = \frac{1}{B} \sum_{u=1}^B \ell_u^\theta = \frac{1}{B} \sum_{u=1}^B \frac{1}{L}\sum_{j=1}^L \ell_{u,j}^\theta \label{eq:loss-erm}
\end{align}

\subsection{Cost-sensitive losses}
As simple baselines, we use importance-weighted losses based on item frequency (CB) and group frequency (IPW) \cite{IW-convergence-mohri,adapative-IW-neurips}. These methods have previously been used as effective baselines for improving group metrics in two-tower recommenders \cite{GoogleDRO10.1145/3485447.3512255}. We also propose a log-weight heuristic as a form of soft-clipping, which often outperforms raw weights in our experiments; these methods are referred to as CBlog and IPWlog, respectively.

\subsubsection{Class balanced (CB) loss:}
Here we weight the loss $\ell^\theta_{u,j}$ inversely proportional to the frequency of the target item $h_u^j$:
\begin{align}
Loss_{CB}^\theta=\frac{1}{B} \sum_{u=1}^B \frac{1}{L} \sum_{j=1}^L w(h_u^j) \cdot \ell^\theta_{u,j}\label{eq:loss-cb}
\end{align}
where the weight $w(h_u^j)=\frac{\sum_{i\in I} f(i)}{f(h_u^j)}$, $f(h_u^j)$ is the number of times item $h_u^j$, appears in the training sequences. CBlog, the variant with log weights replaces $w(h_u^j)$ by $\log(w(h_u^j))$.
\subsubsection{Inverse propensity weighted (IPW) loss:}
Here we weight the loss for user $u$ inversely proportional to the size of the group($g_u$) it belongs to:
\begin{align}
    Loss_{IPW}^\theta &= \frac{1}{B} \sum_{u=1}^B w(g_u) \cdot \ell_u^\theta  \label{eq:IPW}
\end{align}
where the weight $w(g_u) = \frac{\sum_{g \in G} f(g)}{f(g_u)}$, $f(g_u)$ is the number of users belonging to group $g_u$ in the training data. Note that for the IPW loss \eqref{eq:IPW} each user must belong to a single group, and it cannot handle intersecting groups.

\subsubsection{Distributionally robust optimization (DRO):}
The main idea for DRO methods is to minimize the expected loss over the worst case distribution which lies within some distance of the empirical training distribution. In this paper, we consider three different DRO training methods. The following contains a brief description, for further details we point the reader to \cite{cvarpaper} for Conditional Value at Risk DRO, and \cite{SagawagDRO,GoogleDRO10.1145/3485447.3512255} for Group and Streaming DRO respectively.

\subsubsection{Conditional Value at Risk (CVaR) DRO:}
The CVaR loss \cite{cvarpaper,zhai2021boosted} at level $\alpha \in (0,1]$ for a batch of B training examples
\begin{align}
    Loss_{CVaR}^\theta(\alpha) &=  \sup_{q \in \Delta^B}\left\{\sum_{u=1}^B q_u \cdot \ell_u^\theta ~\vert~ \norm{q}_\infty \leq \frac{1}{\alpha B} \right\}, \label{eq:loss-cvar}
\end{align}
where $\Delta^B$ is the probability simplex \footnote{The probability simplex $\Delta^B = \{x \in \R^B | \sum_{k=1}^B x_k =1, x_k \geq 0 ~\forall k\}$} in $\R^B$. The CVaR loss measures how well a model performs over the worst $\alpha$ fraction of the batch. For e.g., if $m = \alpha B$ is an integer then the CVaR loss is the average loss over the $m$ samples that incur the highest losses. In practice, we treat $\alpha$ as a hyperparameter. A key observation is that the loss \eqref{eq:loss-cvar} does not use group memberships. Thus it does not require groups to be defined upfront, and can directly handle users belonging to multiple groups. 

\subsubsection{Group and Streaming DRO:}

Both Group and Streaming DRO build uncertainty sets using group annotations and aim to minimize the maximum loss for a user-group. They maintain a discrete distribution $\omega$ over non-intersecting user groups, with $\sum_{g \in G} \omega_g = 1$, this distribution is first updated using exponentiated gradient ascent with a step-size $\eta$ \cite{SagawagDRO}, which we treat as a hyperparameter in practice. The model parameters $\theta$ are then updated using the loss in \eqref{eq:gdro-desc} using a first order optimizer. 
For streaming DRO, given $L_g^{t}$ the batch loss for group $g$, a streaming estimate \footnote{The streaming estimate $\tilde{L}_g^{t} = (1-\beta) L_g^{t-1} + \beta L_g^{t}$}  $\tilde{L}_g^{t}$  is computed for updating the distribution $\omega$ over the groups \cite{GoogleDRO10.1145/3485447.3512255}. Group DRO directly uses $L_g^{t}$ to update $\omega$. The loss is given by
\begin{align}
Loss_{g/sdro}^\theta &= \sum_{g \in G} \omega_g^t \cdot L_g^{t}. \label{eq:gdro-desc}
\end{align}

\subsubsection{Loss with intersecting groups:}
Note that ERM, CB, and CVaR do not require group memberships and thus generalize to intersecting groups. In contrast, GDRO, SDRO and IPW use group annotations in the loss and are limited to single-group membership per user; we discuss this further in Section \ref{sec:exp-intersecting-groups}

\section{Experiments}
\label{sec:experiments-intro}
In this section, we evaluate the training methods from Section \ref{sec:seqrec+methods} in two scenarios:
\begin{enumerate}
    \item \textit{Single group:} 
     Each user belongs to a single group and we evaluate the training methods across a range group sizes, from balanced to imbalanced.
    \item \textit{Intersecting groups:} 
    Each user belongs to intersecting groups, one popularity based and another based on sequence length. We compare the training methods and analyse the sensitivity of the results for group specific methods.
\end{enumerate}
In Section \ref{sec:Exp-setup} we cover the experimental setup while Section \ref{sec:exp-groupsize-effect} and \ref{sec:exp-intersecting-groups} contain the training method comparisons with single and intersecting groups respectively.

\subsection{Experimental setup}
\label{sec:Exp-setup}
\paragraph{Datasets and preprocessing:} We experiment with (i) Retailrocket \cite{Retailrocketdataset}, an ecommerce dataset from which we use the user views data, and (ii) Movielens-1M \cite{Movielensdataset}, a popular movie dataset. We apply an iterative core-$5$ filtering, ensuring that each user and item has at least $5$ interactions. Table \ref{tab:dataset-stats} summarises the total number of users, items, and interactions.
\paragraph{User groups:} We define two subgroups, G\textsubscript{pop} $=\{\text{niche, diverse, popular}\}$ and  G\textsubscript{seq}$ =\{\text{short, medium, long}\}$. Users belongs to one group from each set. 

\begin{enumerate}
\item G\textsubscript{pop}: Users with extensive interactions tend to consume more long-tail items \cite{Popular-items-Abdollahpuri}, so we define groups based on preferences for these items as in \cite{GoogleDRO10.1145/3485447.3512255}. Specifically, users are grouped by $r_u$, the ratio of popular\footnote{Popular items are those that fall in the top 0.2 quantile of user interactions \cite{Popular-items-Abdollahpuri}} items in their sequence and annotated as niche, diverse, or popular if $r_u$ falls in the bottom, middle, or top quantiles.
\item G\textsubscript{seq}: Sequence length groups are inspired by cold-start literature \cite{volkovs2017dropoutnet,Netflix-coldstart}, where users with different interaction lengths receive recommendations of varying quality. Users are categorized as short, medium, or long based on whether their sequence length falls in the bottom, middle, or top quantiles.
\end{enumerate}
We compare training methods across different group sizes, using three quantile splits that result in balanced (33\% each), semi-balanced (20\%, 60\%, 20\%), and imbalanced (10\%, 80\%, 10\%) groups. Consider the first cell in Table \ref{tab:group-percentages} : if the bottom, middle and top quantile sizes for $r_u$ are 40,16 and 44\% respectively then the training data has an equal proportion of niche, diverse, and popular users.

\paragraph{Backbone:}
We use SASRec, a transformer decoder model, but in contrast to the original SASRec \cite{SasrecOG} which is trained as a binary classification task with negative sampling, we perform multi-class classification with a softmax over the full item corpus followed by the cross entropy loss. Many recent papers \cite{gSAS,dross,acceralated-softmax} show that SASRec trained this way achieves state-of-the-art accuracy.

\paragraph{Evaluation:}
    As is typical in sequential recommenders \cite{SasrecOG,Bert4RecOG} we evaluate using a leave-one-out data split: we hold out the last item in each user's item views sequence for testing while the second to last item is used for validation. We perform a grid search to identify the best architectural parameters for both datasets; further details are provided in Appendix \ref{app:backbone-search}. Following the recommendations in \cite{SampledMetrics,samplingmetrics2}, we avoid negative sampling when measuring model metrics.

\paragraph{Training:}
After identifying the best architectural parameters and convergence epochs, we train SASRec using this fixed epoch budget across the training methods.
Only the DRO-specific \footnote{exponentiated gradient step size $\eta$ for GDRO/SDRO and CVaR level $\alpha$} parameters are tuned, as in \cite{GoogleDRO10.1145/3485447.3512255}, and model selection is based on the highest overall NDCG@20 on the validation set. Using a fixed epoch budget and architecture ensures a fair comparison across methods.

\subsection{Users in a single group}
\label{sec:exp-groupsize-effect}
Previous work \cite{GoogleDRO10.1145/3485447.3512255} defined user groups using a single choice of quantiles then evaluated training methods. However, the performance of the training methods could vary with group size. In this study, we perform a detailed evaluation of the training methods across different group sizes (described in Sec \ref{sec:Exp-setup}, \textit{User groups}).

\begin{table}[!h]
\centering
\begin{subtable}{0.72\textwidth}
    \centering
    \resizebox{\textwidth}{!}{
    \begin{tabular}{lc:c:c|c:c:c}
    \toprule
    Dataset split & \multicolumn{3}{c}{Popularity groups} & \multicolumn{3}{c}{Sequence length groups} \\
    \cline{1-7}
    & $G_{pop33}$ & $G_{pop2060}$ & $G_{pop1080}$ & $G_{seq33}$ & $G_{seq2060}$ & $G_{seq1080}$ \\
    RR dsplit & (33,33,33) & (20,60,20) & (10,80,10) & (33,33,33) & (20,60,20) & (10,80,10) \\
    RR usplit & (40,16,44) & (27,45,28) & (15,70,15) & (75,23,3) & (55,44,0.43) & (34,66,0.09) \\
    \hline
    ML dsplit & (33,33,33) & (20,60,20) & (10,80,10) & (33,33,33) & (20,60,20) & (10,80,10) \\
    ML usplit & (18,28,54) & (10,52,38) & (5,71,23) & (73,19,8) & (59,37,4) & (42,57,2) \\
    \bottomrule
    \end{tabular}
    }
    \caption{\scriptsize{Group sizes in the training data(dsplit) and among the users(usplit) in the Retailrocket views and Movielens-1M datasets. For popularity groups the tuple represents (niche,diverse,popular), for sequence length groups it denotes (short,medium,long)} \label{tab:gsplit}}
    \label{tab:group-percentages}
\end{subtable}
\begin{subtable}{0.27\textwidth}
    \centering
    \resizebox{\textwidth}{!}{
    \begin{tabular}{lccc}
    \toprule
    Dataset & Users & Items & Interactions \\
    \midrule
    RR  & 22178 & 17803 & 364943 \\
    ML1M  & 6040  & 3416  & 999611 \\
    \bottomrule
    \end{tabular}
    }
    \caption{\scriptsize{Dataset statistics}}
    \label{tab:dataset-stats}
\end{subtable}
\caption{(a) We construct three different group sizes subscripted by {\scriptsize (33, 2060, 1080)} for both popularity and sequence length groups denoting balanced, semi-balanced and imbalanced sizes. (b) Number of users, items and interactions}
\label{tab:dinfo}
\end{table}

\begin{table}[!h]
    \centering
    \resizebox{\textwidth}{!}{
    \begin{tabular}{l|llll|llll|llll}
    \toprule
        Method & \multicolumn{4}{c|}{$G_{pop33}$} & \multicolumn{4}{c|}{$G_{pop2060}$} & \multicolumn{4}{c}{$G_{pop1080}$} \\
        & Niche & Diverse & Popular & Overall & Niche & Diverse & Popular & Overall & Niche & Diverse & Popular & Overall \\
        \midrule
        ERM & 0.214 & 0.210 & 0.240 & 0.225 & 0.216 & 0.206 & 0.262 & 0.224 & 0.232 & 0.218 & \underline{0.268} & 0.227 \\
    CB & 0.208\textsuperscript{*} & 0.205 & 0.229\textsuperscript{*} & 0.217\textsuperscript{*} & 0.211\textsuperscript{*} & 0.202\textsuperscript{*} & 0.243\textsuperscript{*} & 0.216\textsuperscript{*} & 0.231 & 0.211\textsuperscript{*} & 0.242\textsuperscript{*} & 0.218\textsuperscript{*} \\
    CBlog & 0.213 & 0.211 & 0.242 & 0.225 & 0.218 & 0.209 & 0.263 & 0.226 & 0.233 & 0.219 & 0.265 & 0.228 \\
    IPW & 0.214 & 0.210 & 0.244\textsuperscript{*} & 0.226 & \underline{0.220}\textsuperscript{*} & 0.201\textsuperscript{*} & 0.266 & 0.224 & \underline{0.238}\textsuperscript{*} & 0.217 & 0.264 & 0.227 \\
    IPWlog & 0.214 & \underline{0.216}\textsuperscript{*} & 0.245\textsuperscript{*} & 0.228\textsuperscript{*} & 0.217 & 0.204 & 0.260 & 0.223 & 0.233 & 0.215 & 0.255\textsuperscript{*} & 0.224\textsuperscript{*} \\
    GDRO & \underline{0.215} & 0.213 & 0.248\textsuperscript{*} & 0.230\textsuperscript{*} & 0.217 & 0.210\textsuperscript{*} & \underline{0.267}\textsuperscript{*} & \underline{0.228}\textsuperscript{*} & 0.234 & \underline{0.220}\textsuperscript{*} & 0.264 & \underline{0.229} \\
    SDRO & 0.214 & 0.215\textsuperscript{*} & \underline{0.250}\textsuperscript{*} & \underline{0.230}\textsuperscript{*} & 0.219 & \underline{0.211}\textsuperscript{*} & 0.259 & 0.227\textsuperscript{*} & 0.233 & 0.220 & 0.266 & 0.228 \\
    CVaR & \textbf{0.225}\textsuperscript{*} & \textbf{0.228}\textsuperscript{*} & \textbf{0.256}\textsuperscript{*} & \textbf{0.239}\textsuperscript{*} & \textbf{0.232}\textsuperscript{*} & \textbf{0.225}\textsuperscript{*} & \textbf{0.269}\textsuperscript{*} & \textbf{0.239}\textsuperscript{*} & \textbf{0.244}\textsuperscript{*} & \textbf{0.229}\textsuperscript{*} & \textbf{0.268} & \textbf{0.237}\textsuperscript{*} \\
    \bottomrule
    \end{tabular}
    }
    \caption{Group and overall NDCG@20 across popularity-based groups on the Retailrocket dataset. $*$ denotes statistically significant difference to ERM ($p< 0.05$, paired T-test). Best in bold, second best is underlined.} 
    \label{tab:RR-poponly}
\end{table}
We first consider the case in which each user only belongs to either the niche, diverse or popular group. Table \ref{tab:RR-poponly} records the NDCG@20 on the Retailrocket views dataset. 
We observe that CVaR achieves the highest NDCG@20 both overall and across user groups, regardless of group size—$G_{pop33}, G_{pop2060}$ and $G_{pop1080}$ (going from balanced to imbalanced).

We now consider the case where each user belongs to the short, medium, or long group. Table \ref{tab:RR-seqonly} shows the NDCG@20 for the Retailrocket views dataset across three group sizes: $G_{seq33}$, $G_{seq2060}$, and $G_{seq1080}$. CVaR continues to demonstrate strong performance. It is important to note that there are only 20 users with long sequence lengths (0.09\% of all users) in the $G_{seq1080}$ split, which explains why the NDCG@20 values in the penultimate column of Table \ref{tab:RR-seqonly} are not statistically significant. We repeat these experiments on the Movielens-1M dataset in Appendix \ref{app:single-ML} with similar insights.

Overall, the summary of this section
is that the DRO methods obtain higher NDCGs \footnote{In a few cases the GDRO and SDRO methods are statistically similar compared to standard training (T-test pvalue $>0.05$), barring these they outperform ERM} than standard training across group sizes. CVaR often surpasses GDRO and SDRO, despite not using group information.

\begin{table}[!h]
        \centering
        \resizebox{\textwidth}{!}{
            \begin{tabular}{l|l l l l|l l l l|l l l l}
            \toprule
            Method & \multicolumn{4}{c|}{$G_{seq33}$} & \multicolumn{4}{c|}{$G_{seq2060}$} & \multicolumn{4}{c}{$G_{seq1080}$} \\
            & Short & Medium & Long & Overall & Short & Medium & Long & Overall & Short & Medium & Long & Overall \\
            \midrule
            ERM & 0.191 & 0.337 & 0.229 & 0.225  & 0.159 & 0.309 & 0.088 & 0.225& 0.127 & \underline{0.282} & 0.013 & \underline{0.229} \\
        CB & 0.188\textsuperscript{*} & 0.323\textsuperscript{*} & 0.218 & 0.219\textsuperscript{*}  & 0.154\textsuperscript{*} & 0.292\textsuperscript{*} & 0.075 & 0.215\textsuperscript{*}& 0.125 & 0.263\textsuperscript{*} & 0.048 & 0.216\textsuperscript{*} \\
        CBlog & 0.190 & 0.332\textsuperscript{*} & 0.228 & 0.224  & 0.157 & 0.307 & 0.090 & 0.223& 0.129 & 0.277\textsuperscript{*} & 0.000 & 0.227\textsuperscript{*} \\
        IPW & 0.194\textsuperscript{*} & 0.340 & 0.230 & 0.228\textsuperscript{*}  & 0.159 & 0.307 & 0.135\textsuperscript{*} & 0.225& 0.120\textsuperscript{*} & 0.272\textsuperscript{*} & \textbf{0.155} & 0.221\textsuperscript{*} \\
        IPWlog & 0.193\textsuperscript{*} & 0.345\textsuperscript{*} & 0.237 & 0.229\textsuperscript{*}  & 0.157 & 0.309 & 0.131 & 0.224& 0.119\textsuperscript{*} & 0.267\textsuperscript{*} & \underline{0.143} & 0.217\textsuperscript{*} \\
        GDRO & \underline{0.201}\textsuperscript{*} & \underline{0.347}\textsuperscript{*} & \underline{0.277}\textsuperscript{*} & 0.237\textsuperscript{*}  & \underline{0.168}\textsuperscript{*} & \underline{0.315}\textsuperscript{*} & 0.064 & \underline{0.233}\textsuperscript{*}& 0.129 & 0.280 & 0.000 & 0.229 \\
        SDRO & 0.201\textsuperscript{*} & \textbf{0.348}\textsuperscript{*} & \textbf{0.279}\textsuperscript{*} & \underline{0.237}\textsuperscript{*}  & 0.165\textsuperscript{*} & 0.313\textsuperscript{*} & \textbf{0.152}\textsuperscript{*} & 0.231\textsuperscript{*}& \underline{0.130}\textsuperscript{*} & 0.279\textsuperscript{*} & 0.011 & 0.229 \\
        CVaR & \textbf{0.207}\textsuperscript{*} & 0.344\textsuperscript{*} & 0.271\textsuperscript{*} & \textbf{0.240}\textsuperscript{*}  & \textbf{0.172}\textsuperscript{*} & \textbf{0.320}\textsuperscript{*} & \underline{0.136}\textsuperscript{*} & \textbf{0.237}\textsuperscript{*}& \textbf{0.137}\textsuperscript{*} & \textbf{0.289}\textsuperscript{*} & 0.083 & \textbf{0.237}\textsuperscript{*} \\
            \bottomrule
            \end{tabular}
        }
    \caption{Group and overall NDCG@20 across sequence-length groups on the Retailrocket dataset. $*$ denotes statistically significant difference to ERM ($p< 0.05$, paired T-test). Best in bold, second best is underlined.}
    \label{tab:RR-seqonly}
\end{table}

\subsection{Users in intersecting groups}
\label{sec:exp-intersecting-groups}
In this section, each user belongs to one of three popularity-based groups and one of three sequence-length groups. We use the $G_{pop33}$ and $G_{seq33}$ splits for popularity and sequence length, respectively, ensuring a balanced training set with 33\% niche, diverse, popular, and 33\% short, medium, long sequence-length users. Table \ref{subtab:RR-popseq} shows the NDCG@20 for the Retailrocket views dataset, where CVaR achieves the highest NDCG@20, except for long-sequence users, where it ranks second. 

\begin{figure}[!h]
        \centering
        \includegraphics[width=\textwidth]{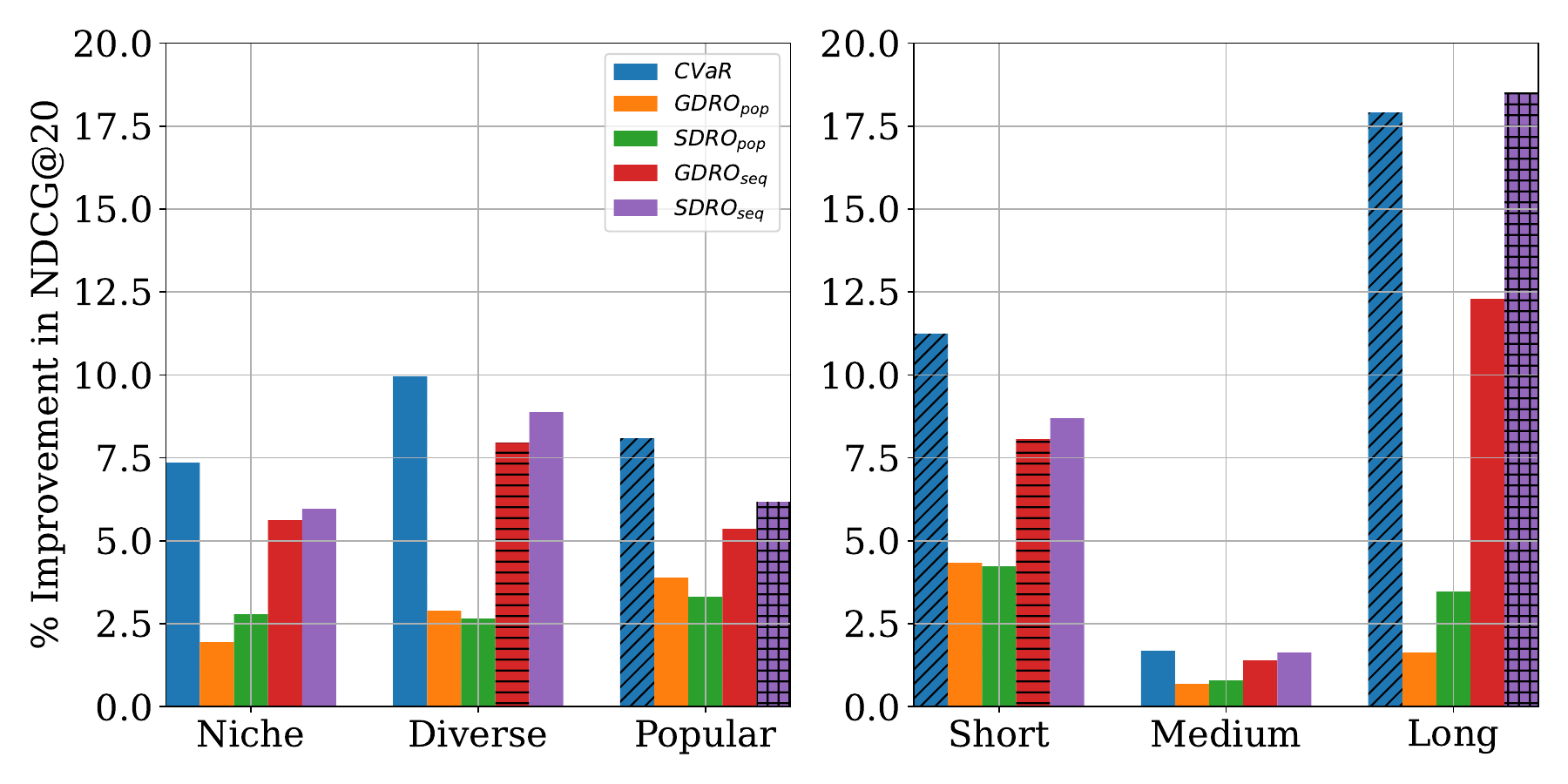}
        \caption{Percentage increase in NDCG@20 using DRO methods relative to standard training on the Retailrocket dataset: (i) increase for niche, diverse and popular groups (ii) increase for short, medium, and long sequence users.}
        \label{fig:RR-popseq}
\end{figure}

Recall that the GDRO, SDRO methods require each user to belong to a single group. To apply these methods, we limit the loss computation to either sequence length or popularity groups. The methods are then referred to with the corresponding subscript such as SDRO\textsubscript{seq} and SDRO\textsubscript{pop}. A key observation for these methods is their sensitivity to the choice of groups. As seen in Figure \ref{fig:RR-popseq} using popularity groups in the loss calculation yields worse relative improvements than using sequence length groups -- GDRO\textsubscript{pop} and SDRO\textsubscript{pop} perform worse than GDRO\textsubscript{seq} and SDRO\textsubscript{seq} respectively. We repeat this experiment on the Movielens-1M dataset in Appendix \ref{app:inter-ML}, observing similar results. 

The conclusion from this section is that training methods reliant on group annotations, like GDRO and SDRO, are sensitive and do not scale to multiple groups. It is impossible to know beforehand which subgroup in the loss will yield the best NDCG scores, and training multiple variants is impractical. While we defined two intersecting subgroups in this paper, real-world scenarios often involve numerous subgroups. As a result, training multiple models with SDRO\textsubscript{pop}, SDRO\textsubscript{seq},  SDRO\textsubscript{subgroup-n} is impractical and even impossible with unknown subgroups.
In contrast, group-agnostic methods like CVaR easily handle multiple subgroups, outperform standard training and even group and streaming DRO.

\begin{table}[!h]
    \centering
        \begin{tabular}{l|l l l l l l |l}
        \toprule
        Method & Niche & Diverse & Popular & Short &  Medium & Long & Overall \\
        \midrule
        ERM & 0.2122 & 0.2080 & 0.2381 & 0.1859 & 0.3418 & 0.2364 & 0.2230 \\
        CB & 0.2115 & 0.2036 & 0.2280\textsuperscript{*} & 0.1861 & 0.3195\textsuperscript{*} & 0.2161\textsuperscript{*} & 0.2175\textsuperscript{*} \\
        CBlog & 0.2154\textsuperscript{*} & 0.2126 & 0.2424\textsuperscript{*} & 0.1927\textsuperscript{*} & 0.3378 & 0.2258 & 0.2269\textsuperscript{*} \\
        CVaR & \textbf{0.2278}\textsuperscript{*} & \textbf{0.2286}\textsuperscript{*} & \textbf{0.2574}\textsuperscript{*} & \textbf{0.2068}\textsuperscript{*} & \textbf{0.3475}\textsuperscript{*} & \underline{0.2787}\textsuperscript{*} & \textbf{0.2409}\textsuperscript{*} \\
        IPW\textsubscript{pop} & 0.2137 & 0.2102 & 0.2432\textsuperscript{*} & 0.1905\textsuperscript{*} & 0.3415 & 0.2327 & 0.2262\textsuperscript{*} \\
        IPWlog\textsubscript{pop} & 0.2139 & 0.2093 & 0.2401 & 0.1873 & 0.3444 & 0.2400 & 0.2247 \\
        GDRO\textsubscript{pop} & 0.2163\textsuperscript{*} & 0.2139\textsuperscript{*} & 0.2473\textsuperscript{*} & 0.1940\textsuperscript{*} & 0.3441 & 0.2402 & 0.2296\textsuperscript{*} \\
        SDRO\textsubscript{pop} & 0.2181\textsuperscript{*} & 0.2135\textsuperscript{*} & 0.2459\textsuperscript{*} & 0.1938\textsuperscript{*} & 0.3445 & 0.2446 & 0.2297\textsuperscript{*} \\
        IPW\textsubscript{seq} & 0.2132 & 0.2098 & 0.2445\textsuperscript{*} & 0.1912\textsuperscript{*} & 0.3401 & 0.2352 & 0.2264\textsuperscript{*} \\
        IPWlog\textsubscript{seq} & 0.2157\textsuperscript{*} & 0.2115 & 0.2455\textsuperscript{*} & 0.1929\textsuperscript{*} & 0.3423 & 0.2312 & 0.2282\textsuperscript{*} \\
        GDRO\textsubscript{seq} & 0.2242\textsuperscript{*} & 0.2245\textsuperscript{*} & 0.2508\textsuperscript{*} & 0.2009\textsuperscript{*} & 0.3465\textsuperscript{*} & 0.2654\textsuperscript{*} & 0.2360\textsuperscript{*} \\
        SDRO\textsubscript{seq} & \underline{0.2249}\textsuperscript{*} & \underline{0.2264}\textsuperscript{*} & \underline{0.2528}\textsuperscript{*} & \underline{0.2021}\textsuperscript{*} & \underline{0.3474}\textsuperscript{*} & \textbf{0.2801}\textsuperscript{*} & \underline{0.2374}\textsuperscript{*} \\
        \bottomrule
        \end{tabular}
        \caption{Group and overall NDCG@20 with users in intersecting groups;       Retailrocket dataset, $*$ denotes statistically significant difference to ERM ($p< 0.05$, paired T-test). Best in bold, second best is underlined.}
        \label{subtab:RR-popseq}
\end{table}

\section{Conclusion}
\label{sec:conclusion}
In this paper, we demonstrated the effectiveness of Conditional Value at Risk DRO in improving both group and overall NDCG scores in sequential recommenders like SASRec. CVaR consistently outperforms Group DRO  and Streaming DRO across various group sizes and in scenarios with intersecting user groups. Based on our experiments, we suggest that practitioners prioritize group-agnostic methods like CVaR over group-dependent approaches, as CVaR not only scales easily to multiple groups but also performs better.
Although we have restricted attention to group attributes that can be inferred from interaction history, such as mainstream inclination and sequence lengths, we note that important user attributes  such as age and ethnicity will commonly be protected and unknown. This effectively renders group-dependent methods inapplicable and strengthens our argument that practitioners should prioritise methods like CVaR.

\bibliography{refs}
\bibliographystyle{plainnat}

\appendix

\section{Experiments on the Movielens-1M dataset}
\label{app:ml1m-exp}
\subsection{Users in a single group (Movielens-1M)}
We repeat the experiments for users in a single group (Section \ref{sec:exp-groupsize-effect}) on the Movielens-1M dataset. Table \ref{tab:ML1m-pop} and Table \ref{tab:ML1m-seq} record the NDCG@20 for users in popularity-based groups and sequence length groups respectively. Here too we observe that the DRO methods obtain higher NDCGs than standard training across group sizes, with CVaR often surpassing GDRO and SDRO. 
\label{app:single-ML}
\begin{table}[!h]
\centering
\begin{subtable}{\textwidth}
\centering
    \resizebox{\textwidth}{!}{
    \begin{tabular}{l|l l l l|l l l l|l l l l}
    \toprule
    Method & \multicolumn{4}{c|}{$G_{pop33}$} & \multicolumn{4}{c|}{$G_{pop2060}$} & \multicolumn{4}{c}{$G_{pop1080}$} \\
    & Niche & Diverse & Popular & Overall & Niche & Diverse & Popular & Overall & Niche & Diverse & Popular & Overall \\
    \midrule
    ERM & 0.189 & 0.179 & 0.208 & 0.196 & \textbf{0.217} & 0.183 & 0.213 & \underline{0.198} & 0.229 & 0.190 & 0.205 & 0.196 \\
CB & 0.123\textsuperscript{*} & 0.109\textsuperscript{*} & 0.126\textsuperscript{*} & 0.121\textsuperscript{*} & 0.137\textsuperscript{*} & 0.113\textsuperscript{*} & 0.127\textsuperscript{*} & 0.121\textsuperscript{*} & 0.152\textsuperscript{*} & 0.112\textsuperscript{*} & 0.126\textsuperscript{*} & 0.117\textsuperscript{*} \\
CB log & 0.187 & 0.178 & 0.204 & 0.194 & 0.209\textsuperscript{*} & 0.181 & 0.208\textsuperscript{*} & 0.194\textsuperscript{*} & 0.234 & 0.188 & 0.201 & 0.194 \\
IPW & 0.188 & 0.180 & 0.208 & 0.196 & 0.200\textsuperscript{*} & 0.170\textsuperscript{*} & 0.206\textsuperscript{*} & 0.187\textsuperscript{*} & 0.190\textsuperscript{*} & 0.162\textsuperscript{*} & 0.182\textsuperscript{*} & 0.168\textsuperscript{*} \\
IPW log & 0.188 & 0.180 & 0.208 & 0.197 & 0.201\textsuperscript{*} & 0.171\textsuperscript{*} & 0.207\textsuperscript{*} & 0.188\textsuperscript{*} & 0.177\textsuperscript{*} & 0.157\textsuperscript{*} & 0.183\textsuperscript{*} & 0.164\textsuperscript{*} \\
GDRO & \underline{0.194} & \textbf{0.184}\textsuperscript{*} & 0.211 & \underline{0.201}\textsuperscript{*} & 0.210 & \underline{0.184} & 0.212 & 0.197 & 0.226 & 0.194\textsuperscript{*} & 0.204 & 0.198 \\
SDRO & 0.191 & \underline{0.184} & \textbf{0.212}\textsuperscript{*} & 0.201\textsuperscript{*} & 0.214 & 0.182 & \underline{0.213} & 0.197 & \underline{0.237} & \underline{0.196}\textsuperscript{*} & \textbf{0.209} & \underline{0.201}\textsuperscript{*} \\
CVaR & \textbf{0.196} & 0.183 & \underline{0.212}\textsuperscript{*} & \textbf{0.201}\textsuperscript{*} & \underline{0.215} & \textbf{0.188}\textsuperscript{*} & \textbf{0.214} & \textbf{0.201}\textsuperscript{*} & \textbf{0.238} & \textbf{0.196}\textsuperscript{*} & \underline{0.209} & \textbf{0.201}\textsuperscript{*} \\
    \bottomrule
    \end{tabular}
}
    \label{subtab:ML1m-pop}
    \end{subtable}
    \caption{Group and overall NDCG@20 across popularity-based groups on the Movielens-1M dataset. $*$ denotes statistically significant difference to ERM ($p< 0.05$, paired T-test). Best in bold, second best is underlined.}
    \label{tab:ML1m-pop}
\end{table}

\begin{table}[!h]
        \centering
        \resizebox{\textwidth}{!}{
            \begin{tabular}{l|l l l l|l l l l|l l l l}
            \toprule
            Method & \multicolumn{4}{c|}{$G_{seq33}$} & \multicolumn{4}{c|}{$G_{seq2060}$} & \multicolumn{4}{c}{$G_{seq1080}$} \\
            & Short & Medium & Long & Overall & Short & Medium & Long & Overall & Short & Medium & Long & Overall \\
            \midrule
            ERM & 0.220 & 0.139 & 0.132 & 0.198  & 0.227 & 0.156 & 0.131 & 0.197& 0.238 & 0.167 & 0.138 & 0.196 \\
        CB & 0.136\textsuperscript{*} & 0.078\textsuperscript{*} & 0.081\textsuperscript{*} & 0.121\textsuperscript{*}  & 0.138\textsuperscript{*} & 0.090\textsuperscript{*} & 0.079\textsuperscript{*} & 0.118\textsuperscript{*}& 0.149\textsuperscript{*} & 0.102\textsuperscript{*} & 0.081\textsuperscript{*} & 0.121\textsuperscript{*} \\
        CBlog & 0.216\textsuperscript{*} & 0.136 & 0.127 & 0.194\textsuperscript{*}  & 0.224 & 0.153 & 0.126 & 0.194\textsuperscript{*}& 0.237 & 0.164\textsuperscript{*} & 0.123\textsuperscript{*} & 0.194\textsuperscript{*} \\
        IPW & 0.218\textsuperscript{*} & 0.135 & 0.132 & 0.196\textsuperscript{*}  & 0.213\textsuperscript{*} & 0.143\textsuperscript{*} & 0.103\textsuperscript{*} & 0.183\textsuperscript{*}& 0.211\textsuperscript{*} & 0.128\textsuperscript{*} & 0.091\textsuperscript{*} & 0.162\textsuperscript{*} \\
        IPWlog & 0.219 & 0.140 & 0.137 & 0.197  & 0.205\textsuperscript{*} & 0.141\textsuperscript{*} & 0.105\textsuperscript{*} & 0.177\textsuperscript{*}& 0.207\textsuperscript{*} & 0.126\textsuperscript{*} & 0.081\textsuperscript{*} & 0.159\textsuperscript{*} \\
        GDRO & \underline{0.222} & \underline{0.148} & \underline{0.143} & \textbf{0.202}  & \underline{0.235}\textsuperscript{*} & \textbf{0.167}\textsuperscript{*} & \underline{0.147} & \textbf{0.206}\textsuperscript{*}& \underline{0.252}\textsuperscript{*} & \textbf{0.176}\textsuperscript{*} & \underline{0.151} & \textbf{0.207}\textsuperscript{*} \\
        SDRO & 0.221 & \textbf{0.149}\textsuperscript{*} & \textbf{0.145} & \underline{0.202}  & \textbf{0.235}\textsuperscript{*} & \underline{0.164}\textsuperscript{*} & \textbf{0.148} & \underline{0.205}\textsuperscript{*}& \textbf{0.253}\textsuperscript{*} & \underline{0.175}\textsuperscript{*} & 0.145 & \underline{0.207}\textsuperscript{*} \\
        CVaR & \textbf{0.222} & 0.143 & 0.139 & 0.201  & 0.230 & 0.161\textsuperscript{*} & 0.137 & 0.201\textsuperscript{*}& 0.249\textsuperscript{*} & 0.174\textsuperscript{*} & \textbf{0.155} & 0.205\textsuperscript{*} \\
            \bottomrule
            \end{tabular}
        }
    \label{subtab:ML1m-seq}
    \caption{Group and overall NDCG@20 across sequence-length groups on the Movielens-1M dataset. $*$ denotes statistically significant difference to ERM ($p< 0.05$, paired T-test). Best in bold, second best is underlined.}
    \label{tab:ML1m-seq}
\end{table}
\subsection{Users in intersecting groups (Movielens-1M)}
\begin{figure}[!h]
        \centering
        \includegraphics[width=\textwidth]{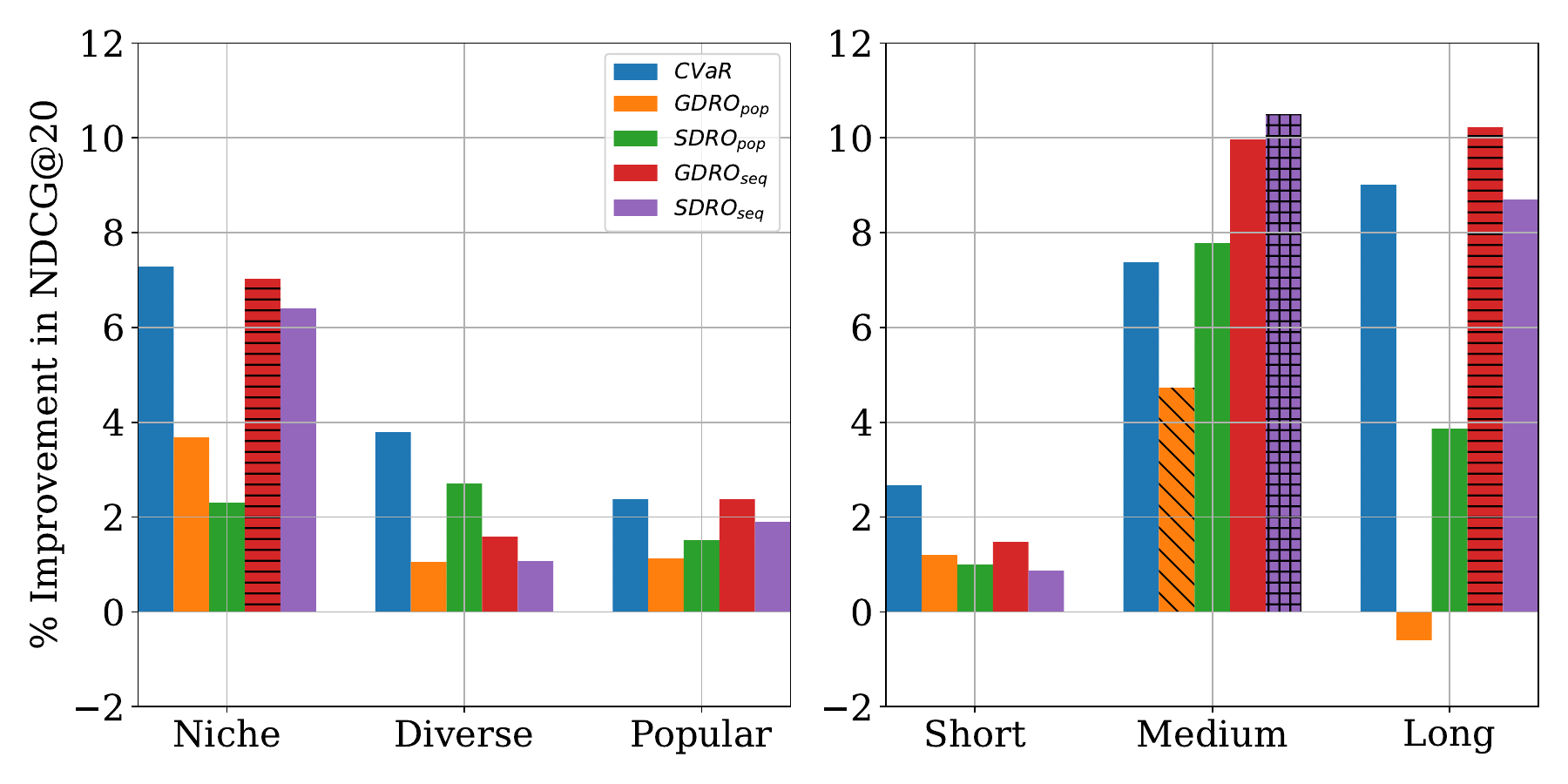}
        \caption{Percentage increase in NDCG@20 using DRO methods relative to standard training on the Movielens-1M dataset: (i) increase for niche, diverse, and popular groups, and (ii) increase for short, medium, and long sequence users.}
        \label{fig:ML1m-popseq}
\end{figure}

We repeat the experiments for users in intersecting groups (Section \ref{sec:exp-intersecting-groups}) on the Movielens-1M dataset. Table \ref{tab:ML1m-popseq} presents the NDCG@20 scores, showing that CVaR achieves the highest NDCG@20 overall and across most user groups. Figure \ref{fig:ML1m-popseq} again illustrates the sensitivity of the GDRO and SDRO results to the choice of groups in the loss computation.
\label{app:inter-ML}
\begin{table}[!h]
        \centering
        \begin{tabular}{l|l l l l l l |l}
            \toprule
            Method & Niche & Diverse & Popular & Short &  Medium & Long & Overall \\
            \midrule
            ERM & 0.1859 & 0.1793 & 0.2092 & 0.2194 & 0.1350 & 0.1314 & 0.1966 \\
        CB & 0.1252\textsuperscript{*} & 0.1099\textsuperscript{*} & 0.1254\textsuperscript{*} & 0.1357\textsuperscript{*} & 0.0792\textsuperscript{*} & 0.0822\textsuperscript{*} & 0.1209\textsuperscript{*} \\
        CBlog & 0.1870 & 0.1781 & 0.2035\textsuperscript{*} & 0.2155\textsuperscript{*} & 0.1354 & 0.1257 & 0.1934\textsuperscript{*} \\
        CVaR & \textbf{0.1994}\textsuperscript{*} & \textbf{0.1861}\textsuperscript{*} & \textbf{0.2142}\textsuperscript{*} & \textbf{0.2252}\textsuperscript{*} & 0.1449\textsuperscript{*} & \underline{0.1433} & \textbf{0.2037}\textsuperscript{*} \\
        IPW\textsubscript{pop} & 0.1875 & 0.1777 & 0.2064\textsuperscript{*} & 0.2171 & 0.1368 & 0.1276 & 0.1949 \\
        IPWlog\textsubscript{pop} & 0.1913\textsuperscript{*} & 0.1796 & 0.2074 & 0.2197 & 0.1338 & 0.1327 & 0.1967 \\
        GDRO\textsubscript{pop} & 0.1927 & 0.1812 & 0.2115 & 0.2220 & 0.1414\textsuperscript{*} & 0.1307 & 0.1996\textsuperscript{*} \\
        SDRO\textsubscript{pop} & 0.1901 & \underline{0.1842} & 0.2124 & 0.2215 & 0.1455\textsuperscript{*} & 0.1365 & 0.2005\textsuperscript{*} \\
        IPW\textsubscript{seq} & 0.1915\textsuperscript{*} & 0.1787 & 0.2088 & 0.2190 & 0.1385 & 0.1351 & 0.1972 \\
        IPWlog\textsubscript{seq} & 0.1921\textsuperscript{*} & 0.1793 & 0.2097 & 0.2200 & 0.1399\textsuperscript{*} & 0.1319 & 0.1980 \\
        GDRO\textsubscript{seq} & \underline{0.1989}\textsuperscript{*} & 0.1821 & \underline{0.2142} & \underline{0.2226} & \underline{0.1484}\textsuperscript{*} & \textbf{0.1449} & \underline{0.2025}\textsuperscript{*} \\
        SDRO\textsubscript{seq} & 0.1977\textsuperscript{*} & 0.1812 & 0.2132 & 0.2213 & \textbf{0.1491}\textsuperscript{*} & 0.1429 & 0.2015\textsuperscript{*} \\
        \bottomrule
        \end{tabular}
        \caption{Group and overall NDCG@20 with users in intersecting groups;       Movielens-1M dataset, $*$ denotes statistically significant difference to ERM ($p< 0.05$, paired T-test). Best in bold, second best is underlined.}
        \label{tab:ML1m-popseq}
\end{table}

\section{Hyperparameters}
\label{app:backbone-search}
\paragraph{Backbone search:}
We conduct a grid search to determine the best architectural parameters for each dataset. Following \cite{dross,RecJPQ,SasrecOG}, we truncate or pad sequences to $L = 200$ most recent interactions, vary embedding dimensions in $\{128, 256\}$, feed-forward dimension in $\{128, 256\}$, number of transformer blocks in $\{1, 2, 3\}$, and the number of attention heads in $\{1, 2, 4\}$. We also test dropout values in $\{0.1, 0.2, 0.5\}$, use the Adam optimizer with a learning rate of $0.001$, and vary batch size $B$ in  $\{128, 256\}$. For each configuration, we run standard training and early stop if validation NDCG@20 does not improve for 200 epochs, each epoch consists of $128$ mini-batches. The best backbone configurations are reported in Table \ref{tab:dataset_hyp}.

\begin{table}[h!]
\centering
\begin{tabular}{c|c|c|c|c|c|c}
\hline
\textbf{Dataset} & \textbf{emb-dim} & \textbf{ff-dim} & \textbf{\#blocks} & \textbf{\#heads} & \textbf{dropout} & \textbf{B} \\ \hline
RR  & 256 & 256 & 3 & 1 & 0.2 & 128                  \\ \hline
ML1M & 256 & 256 & 3 & 1 & 0.5 & 256                 \\ \hline
\end{tabular}
\caption{Best SASRec parameters for the Retailrocket, Movielens-1M datasets}
\label{tab:dataset_hyp}
\end{table}
\paragraph{Computational resources:}
We launch training jobs on Amazon SageMaker using ml.p3.2xlarge instances, each with a Tesla V100 GPU (16GB memory). A single training job takes approximately 8 hours for the Retailrocket dataset and 6 hours for the MovieLens-1M dataset. For GDRO and SDRO, we run five jobs with $\eta$ values from $\{1\text{e}^{-3}, 5\text{e}^{-3}, 1\text{e}^{-2}, 5\text{e}^{-2}, 0.1\}$, and for CVaR, ten jobs with $\alpha$ levels from $\{0.1, 0.2 \ldots 0.9, 0.95\}$. In Section \ref{sec:exp-intersecting-groups}, for users in two intersecting groups, we train two variants for group-specific methods (e.g., GDRO\textsubscript{pop} and GDRO\textsubscript{seq}), launching 10 jobs each for GDRO and SDRO.

\end{document}